\documentclass[a4paper,aps,twocolumn]{revtex4}
\RequirePackage[english]{babel}
\RequirePackage[latin1]{inputenc}
\RequirePackage[T1]{fontenc}
\RequirePackage{mathrsfs}
\RequirePackage{amsmath}
\RequirePackage{amssymb}
\RequirePackage{amsbsy}
\RequirePackage{bm}
\begin{document}
%%%%%%%%%%%%%%%%%%%%%%%%%%%%%%%%%%%%%%%%%%%%%%%%%%%%%%%%%%%%%%%%%%%%%%%%%%%%%%%%%%%%%%%%%%%%%%%%%%%
\title{\bf{Least-Order Torsion-Gravity for Dirac Fields, and\\
their Non-Linearity terms}}
\author{Luca Fabbri}
\affiliation{INFN \& Dipartimento di Fisica, Universit\`{a} di Bologna,\\
Via Irnerio 46, 40126 Bologna, ITALY}
\date{\today}
%%%%%%%%%%%%%%%%%%%%%%%%%%%%%%%%%%%%%%%%%%%%%%%%%%%%%%%%%%%%%%%%%%%%%%%%%%%%%%%%%%%%%%%%%%%%%%%%%%%
\begin{abstract}
We will consider the most general least-order torsional completion of gravity with electrodynamics for the Dirac matter fields, and we study the effects that the torsion-spin coupling will have in inducing self-interactions among the fermion fields themselves; we will see that such self-interactions of fermions have effects analogous to those of the field-quantization prescription, and we will study the way in which they can give rise to matter distributions that are localized in a compact region and stable under the influence of perturbations.
\end{abstract}
%%%%%%%%%%%%%%%%%%%%%%%%%%%%%%%%%%%%%%%%%%%%%%%%%%%%%%%%%%%%%%%%%%%%%%%%%%%%%%%%%%%%%%%%%%%%%%%%%%%
\maketitle
%%%%%%%%%%%%%%%%%%%%%%%%%%%%%%%%%%%%%%%%%%%%%%%%%%%%%%%%%%%%%%%%%%%%%%%%%%%%%%%%%%%%%%%%%%%%%%%%%%%
\section*{Introduction}
In the present paper, we will consider the most general case of least-order derivative torsional completion of gravity with electrodynamics for the $\frac{1}{2}$-spin fermionic Dirac matter fields: the action will be analyzed by obtaining the field equations, in which we will perform the customary treatment separating torsion and having it substituted in terms of the spin of the spinorial matter, with the consequence that there will arise non-linear interactions of the matter fields; these non-linear terms are essentially incompatible with the superposition principle that lies at the basis of the quantum. What this means is that one of the two must be disregarded in order for the other to work properly: the way in which this is usually done is by disregarding torsion, or even more drastically by assuming that the whole gravitational theory needs to be included within an extended frame, while disregarding quantum principles is not even conceived as an alternative worth pursuing. In this paper, we would like to discuss this situation in a somewhat detailed manner.

In order to do so, we have to start from the beginning, and the first thing to do is to briefly recall what is taken as the basic protocol of field quantization: in the usual prescription, after a given theory of fields is developed, the common way of dealing with it is to quantize it, by assuming a plane-wave form for the solution, and after that the solution is found, by expanding the plane waves on a basis of creation/annihilation operators on which commutation relationships are imposed; after this is done, the resulting quantum field theory is capable of achieving predictions of impressive precision, and therefore it is considered to be the most accurate theory ever accomplished in history. With a precision that in some cases might even amount to the twelfth decimal place, it is difficult not to be tempted to consider such a theory the correct description of nature, despite the philosophical warning that no matter how many predictions and however precise they are a theory is never confirmed beyond any reasonable doubt; on the other hand, quantum field theory is widely believed to have given rise to no wrong prediction, and this increases the temptation, despite the philosophical warning that a single wrong prediction is enough to rule out a theory. As a consequence of all this, given that quantum field theory is enjoying a successful situation, why would anyone resist to such temptation, that is why would not anyone embrace quantum field theory entirely, beyond any reasonable doubt?

The answer is that people usually don't, and as a consequence quantum field theory is considered to be the real description of nature; and as another consequence any conflicting theory must be the one with problems, the one that has to be disregarded. But nevertheless, we would like to go a little deeper in discussing this issue.

The general outline of this discussion starts with the following question: can we really assume that predictions up to twelve decimal places are enough, in order to take quantum field theory as the correct description of nature beyond any doubt? Twelve decimal places are an impressive precision, and one may reasonably be tempted to guess that the answer to this question ought be affirmative, but we would like to bring an example in the history of physics in which even greater precisions were not enough to give an affirmative reply: the Bohr atom was the first attempt to employ strings in the history of physics, as in Bohr's picture the electron was considered to be a string encircling the nucleus in such a way that it could have an integer number of oscillations contained within the length of the orbit; after this was considered, Bohr has been able to predict the formula for the energy levels of the electron. This is an important point, because Bohr has not been able to predict the energy levels at the twelve decimal place but with an infinite precision, since what he got was the exact formula, and in this sense any prediction of quantum field theory is infinitely less precise than the energy levels in the atom described by Bohr; nevertheless, nowadays nobody would seriously think that the electron encircles the nucleus in this manner. The reason for which the Bohr atom has been abandoned was that it was improved by the Schr\"{o}dinger equation, which could not only predict the same energy levels, but also the electronic orbitals enveloping the nucleus. However, it is clear that Bohr atom is indeed very predictive but nevertheless wrong, and that this has nothing to do with the improvement given by the Schr\"{o}dinger equation. Great precisions are not enough to consider a theory correct beyond any doubt, and this even if there is no alternative available. As a consequence of this, the philosophical warning must be considered.

A second point we need assess is whether quantum field theory has ever had a wrong prediction: is there a situation in which quantum field theory predicted an effect that was falsified by experiments? Generally, the answer to this question is negative, but here too there is more that has to be said: because in quantum field theory one has to cope with commutation relationships, and placing annihilation operators on the left and creation operators on the right has the effect of producing delta distributions in the energy density of the fields that remain even in the vacuum configuration. When these zero-point energies are integrated over the volume of the universe, they result into a contribution to the cosmological constant that is off the measured value by some one-hundred twenty orders of magnitude. Unfortunately, a custom that happens to be followed in quantum field theory is that a prediction that is off the observed value is not taken as a wrong prediction, but rather it is called anomaly and it is later assumed to be cancelled by using further adjustment and fine-tuning. If wrong predictions are hidden in order to be eventually treated then no theory would ever be falsified. So in this circumstance too the philosophical warning must be taken seriously.

This said, our position is becoming clearer: quantum field theory has reached predictions whose precision is astonishing, but we have seen that other models had reached even larger precisions and yet they have been acknowledged to be wrong, so we cannot consider the present evidence as a confirmation that quantum field theory is correct beyond any doubt; on the other hand, we have also seen that quantum field theory is the protagonist of wrong predictions, and we must consider such evidence as proof that quantum field theory is incorrect.

Another consideration supporting this view is on the compatibility between torsion-gravity for Dirac fields and quantization applied to these fields: as we have said in introducing the theory, this theory is based on the assumptions of having least-order derivative coupling, and the requirement of having the most general of such models inevitably leads to the torsion-gravitational matter field theory, in which torsionally-induced non-linear interactions necessarily arise; on the other hand, quantum field theory is based on the assumption of having renormalizable actions plus the hypothesis of linearity plus the additional hypothesis of field-quantization. If the assumption of least-order derivative is as strong as renormalizability, certainly the hypotheses of linearity and field quantization are supplementary hypotheses: when two sets of postulates are confronted on theoretical grounds, the model that has more or more arbitrary axioms is the one that tends to be disfavoured. So again, philosophical warnings come to bring critical arguments in this balance.

Therefore, in what follows, our position will be that of renouncing to the possibility to have any form of quantization of fields: this assumption may be considered as extreme but nevertheless it is equivalent to implementing the normal-ordering procedure \cite{p-s}; this procedure in quantum field theory is usually implemented whenever it is convenient but not in general because implementing it on the commutation relationships would merely mean that there would be no commutation relationships, and therefore there would remain no quantization in quantum field theory. But here such a prescription can and will be implemented in general, in all possible circumstances.

In this paper, therefore, we will consider no quantization of any field, whatsoever; as a consequence of this assumption, we will have to find a way in which to recover the effects due to such a prescription in terms of something different, and because quantum superposition is incompatible with non-linear potentials, it would be just right if quantum effects might be recovered precisely in terms of those non-linearity terms: further properties might be obtained if exact solutions were to be found.

This situation is rather curious, and therefore we will spend some word to comment about it eventually.
%%%%%%%%%%%%%%%%%%%%%%%%%%%%%%%%%%%%%%%%%%%%%%%%%%%%%%%%%%%%%%%%%%%%%%%%%%%%%%%%%%%%%%%%%%%%%%%%%%%
%%%%%%%%%%%%%%%%%%%%%%%%%%%%%%%%%%%%%%%%%%%%%%%%%%%%%%%%%%%%%%%%%%%%%%%%%%%%%%%%%%%%%%%%%%%%%%%%%%%
\section{Least-Order Torsion-Gravity for Dirac Fields}
%%%%%%%%%%%%%%%%%%%%%%%%%%%%%%%%%%%%%%%%%%%%%%%%%%%%%%%%%%%%%%%%%%%%%%%%%%%%%%%%%%%%%%%%%%%%%%%%%%%
\subsection{Geometric and Kinematic Quantities}
To begin, we shall introduce the geometric and kinematic quantities we will employ in the present work.

All along the present paper, the geometry we will employ is based on a $(1\!+\!3)$-dimensional spacetime and it will be a Riemann-Cartan geometry: the process of raising/lowering indices in tensors is possible by introducing two tensors $g^{\mu\nu}$ and $g_{\mu\nu}$ which will be considered as the spacetime metric tensors; differential properties must preserve covariance, and thus they are given by covariant derivatives $D_{\mu}$ which can be defined after introducing the connection $\Gamma^{\alpha}_{\mu\nu}$ whose antisymmetric part in the two lower indices $\Gamma^{\alpha}_{\mu\nu}\!\!\!-\!\!\Gamma^{\alpha}_{\nu\mu}\!\!=\!\!\! Q^{\alpha}_{\phantom{\alpha}\mu\nu}$ is nevertheless a tensor, which is called Cartan torsion tensor. We demand that the following condition $D_{\mu}g_{\alpha\beta}\!=\!0$ holds, and this condition is called metric-compatibility condition; the torsion tensor will be assumed to be completely antisymmetric, for the reason that follows: metric-compatibility ensures the Lorentz structure is preserved \cite{h,h-h-k-n}, and as a consequence the system of coordinates in which locally the metric is Minkowskian and the connection vanishes coincide, while a completely antisymmetric torsion ensures that there is a single symmetric part in the connection
\begin{eqnarray}
&\Gamma^{\alpha}_{\beta\mu}\!=\!\frac{1}{2}g^{\alpha\rho}(\partial_{\beta}g_{\mu\rho}
\!+\!\partial_{\mu}g_{\beta\rho}\!-\!\partial_{\rho}g_{\mu\beta}\!+\!Q_{\rho\beta\mu})
\end{eqnarray}
as the most general decomposition. So the conditions of metric-compatibility and completely antisymmetric torsion together ensure that we can always find a coordinate system where locally both the metric is Minkowskian and the symmetric part of the connection vanishes. This fact allows the implementation of the light-cone structure and the local free-fall as discussed in \cite{a-l, m-l}. Therefore, causality and the principle of equivalence can be mathematically implemented \cite{xy}. And gravitation is geometrized.

From the metric we can define the completely antisymmetric tensor of Levi-Civita $\varepsilon_{\rho\mu\nu\alpha}$ as usual; we also introduce the covariant derivative $\nabla_{\mu}$ defined in terms of the connection $\Lambda^{\alpha}_{\mu\nu}$ which is called Levi-Civita connection, and it is the simplest connection in the sense that it is symmetric in the two lower indices and it is written in terms of the metric alone. We have that the metric-compatibility $\nabla_{\mu}g_{\alpha\beta}\!=\!0$ holds identically; the completely antisymmetric torsion can be written according to the following expression $Q_{\alpha\mu\nu}\!=\! \varepsilon_{\alpha\mu\nu\sigma} W^{\sigma}$ in terms of the dual of the axial vector $W_{\alpha}$ as it is clear: then we have that the covariant derivatives of the Levi-Civita tensor vanish identically, and the above connection is written as
\begin{eqnarray}
&\Gamma^{\alpha}_{\beta\mu}\!=\!\Lambda^{\alpha}_{\beta\mu}
\!+\!\frac{1}{2}g^{\alpha\rho}\varepsilon_{\rho\beta\mu\sigma}W^{\sigma}
\end{eqnarray}
as an equivalent decomposition, holding in general.

This formalism with Greek indices is called coordinate formalism, and it is possible to introduce an equivalent but different formalism with Latin indices called Lorentz formalism, with the advantage that in it, the most general coordinate transformation is converted without loss of generality into the special Lorentz transformation, whose specific form may be explicited and therefore given in terms of some different representation: in Lorentz formalism, the metric is decomposed according to the following $g^{\alpha\nu}\!=\!e^{\alpha}_{p}e^{\nu}_{i} \eta^{pi}$ and $g_{\alpha\nu}\!=\!e_{\alpha}^{p}e_{\nu}^{i} \eta_{pi}$ in terms of the tetrad basis given by $e_{\alpha}^{i}$ and the dual $e^{\alpha}_{i}$ and in terms of the constant metric $\eta_{ij}$ defined to have Minkowskian structure, the one which will have to be preserved by the Lorentz transformation; differential properties preserving also this type of covariance are defined analogously in terms of the covariant derivative $D_{\mu}$ defined after introducing the spin-connection $\omega^{i}_{\phantom{i}j\nu}$ and from which we can define no torsion. Nonetheless, we have that metric-compatibilities $D_{\mu}e_{i}^{\nu}\!=\!0$ and $D_{\mu}\eta_{ij}\!=\!0$ hold; it is however possible by employing such metric-compatibilities to convert the above torsion into this formalism as
\begin{eqnarray}
&-Q^{k}_{\alpha\rho}=\partial_{\alpha}e_{\rho}^{k}-\partial_{\rho}e_{\alpha}^{k}
+\omega^{k}_{\phantom{i}p\alpha}e_{\rho}^{p}-\omega^{k}_{\phantom{i}p\rho}e_{\alpha}^{p}
\end{eqnarray} 
as it is easy to check directly: we notice that the metric-compatibility conditions applied on the Minkowskian metric implies that $\omega^{ip}_{\phantom{ip}\alpha} \!=\! -\omega^{pi}_{\phantom{pi}\alpha}$ spelling the antisymmetry in the two Lorentz indices of the spin-connection itself and therefore ensuring local Lorentz structure to be preserved, while the metric-compatibility applied on the tetrad implies that the spin-connection be given by
\begin{eqnarray}
&\omega^{i}_{\phantom{i}p\alpha}\!=\!
e^{i}_{\sigma}e^{\rho}_{p}(e^{k}_{\rho}\partial_{\alpha}e^{\sigma}_{k}
\!+\!\Gamma^{\sigma}_{\rho\alpha})
\end{eqnarray} 
in the most general case possible. Thus these two metric-compatibility conditions are what ensures that the coordinate formalism and the Lorentz formalism are equivalent, and that the latter is better equipped to incorporate local Lorentz structures: this will be important when dealing with spinors. As it was anticipated, with the Lorentz formalism we look for other representations.

Since one of the possible different representations is the complex representation, it is useful to introduce also the geometry of complex fields, in which the transformation is given by a complex unitary phase: gauge covariant derivatives $D_{\mu}$ are defined after introducing the gauge connection $A_{\nu}$ as it is well known. We will see why it is important to introduce such an abelian gauge structure.

We will introduce the Lorentz group in complex representations, and we will restrict ourselves to the least-spin given by the $\frac{1}{2}$-spin representation only: such a representation can be achieved through the introduction of the gamma matrices $\{\boldsymbol{\gamma}_{a},\boldsymbol{\gamma}_{b}\}\!=\!2\boldsymbol{\mathbb{I}}\eta_{ab}$ from which one may define the matrices $\boldsymbol{\sigma}_{ab}\!=\!\frac{1}{4}
[\boldsymbol{\gamma}_{a},\boldsymbol{\gamma}_{b}]$ as the infinitesimal generators of the Lorentz transformation that is written in the complex $\frac{1}{2}$-spin representation, and these matrices also verify the relations $\{\boldsymbol{\gamma}_{a},\boldsymbol{\sigma}_{bc}\}\!=\!
i\varepsilon_{abcd} \boldsymbol{\pi}\boldsymbol{\gamma}^{d}$ implicitly defining the $\boldsymbol{\pi}$ matrix which will be used to define the left-handed and the right-handed projectors; differential properties are given by the most general spinorial covariant derivative $\boldsymbol{D}_{\mu}$ defined upon introduction of the most general spinorial connection $\boldsymbol{\Omega}_{\mu}$ and there is no torsion defined in its terms. Conditions $\boldsymbol{D}_{\mu}\boldsymbol{\gamma}_{a}\!=\!0$ are valid automatically as identities; notice that in this case it is not even possible to express the already known torsion tensor in spinorial form: from the conditions of metric-compatibility it is possible to see that the spinorial connection can be written according to the form
\begin{eqnarray}
&\boldsymbol{\Omega}_{\rho}
=\frac{1}{2}\omega^{ij}_{\phantom{ij}\rho}\boldsymbol{\sigma}_{ij}
\!+\!iqA_{\rho}\mathbb{I}
\end{eqnarray}
as a general decomposition. Thus we finally see that the antisymmetry in the two Lorentz indices of the spin-connection is mirrored by the antisymmetry in the two indices of the generators of the Lorentz transformation, and this is what ensures local Lorentz structure to be preserved in this formalism: the most general spinorial connection is not however exhausted by the Lorentz group since there is still room for an abelian field which may now be identified with the abelian gauge field above in terms of its charge $q$ as usual. This fact has the most intriguing consequence that dealing with spinor fields ultimately means to work with a spinorial connection whose most general form can exactly be decomposed in terms of a torsion-gravity plus an abelian gauge part, so that both gravitational and the electrodynamic information are altogether stored within the most general spinorial connection we may have. As a consequence, this fact represents some sort of unifying description of the geometrical interactions, a description that is furnished in the case of spinorial matter fields in a natural way.

So far we have introduced a very compact but quite self-contained overview of the general formalism we will employ in the present paper, a formalism that was based on the notion of covariant derivative, but of course it is possible to see what additional structures may be defined if we go to a higher-order derivative, studying the commutator of two covariant derivatives: the first quantity we have to introduce, and possibly the most fundamental, is the curvature tensor of the connection, defined as 
\begin{eqnarray}
&G^{\rho}_{\phantom{\rho}\xi\mu\nu}
=\partial_{\mu}\Gamma^{\rho}_{\xi\nu}-\partial_{\nu}\Gamma^{\rho}_{\xi\mu}
+\Gamma^{\rho}_{\sigma\mu}\Gamma^{\sigma}_{\xi\nu}
-\Gamma^{\rho}_{\sigma\nu}\Gamma^{\sigma}_{\xi\mu}
\end{eqnarray}
antisymmetric in both first and second pair of indices and such that it verifies the cyclic permutation property
\begin{eqnarray}
\nonumber
&D_{\kappa}Q^{\rho}_{\phantom{\rho}\mu \nu}
\!+\!D_{\nu}Q^{\rho}_{\phantom{\rho} \kappa \mu}
\!+\!D_{\mu}Q^{\rho}_{\phantom{\rho} \nu \kappa}\!+\!\\
\nonumber
&+Q^{\pi}_{\phantom{\pi} \nu \kappa}Q^{\rho}_{\phantom{\rho}\mu \pi}
\!+Q^{\pi}_{\phantom{\pi}\mu \nu}Q^{\rho}_{\phantom{\rho}\kappa \pi}
\!+Q^{\pi}_{\phantom{\pi}\kappa \mu}Q^{\rho}_{\phantom{\rho}\nu \pi}-\\
&-G^{\rho}_{\phantom{\rho}\kappa \nu \mu}
\!-\!G^{\rho}_{\phantom{\rho}\mu \kappa \nu}
\!-\!G^{\rho}_{\phantom{\rho}\nu \mu \kappa}\equiv0
\label{tor}
\end{eqnarray}
and because of these antisymmetry properties, the curvature has a single contraction given by $G^{\rho}_{\phantom{\rho}\mu\rho\nu}\!=\!G_{\mu\nu}$ with contraction $G_{\eta\nu}g^{\eta\nu}\!=\!G$ called Ricci tensor and scalar respectively; the curvature comes along with its torsionless counterpart given in terms of the analogous expression
\begin{eqnarray}
&R^{\rho}_{\phantom{\rho}\xi\mu\nu}
=\partial_{\mu}\Lambda^{\rho}_{\xi\nu}-\partial_{\nu}\Lambda^{\rho}_{\xi\mu}
+\Lambda^{\rho}_{\sigma\mu}\Lambda^{\sigma}_{\xi\nu}
-\Lambda^{\rho}_{\sigma\nu}\Lambda^{\sigma}_{\xi\mu}
\end{eqnarray}
antisymmetric in both first and second pair of indices, and such that it verifies the cyclic permutation property given by $R^{\rho}_{\phantom{\rho}\kappa \nu \mu}\!+\!
R^{\rho}_{\phantom{\rho}\mu \kappa \nu}\!+\!R^{\rho}_{\phantom{\rho}\nu \mu \kappa}\equiv0$ and in the same way we have a single contraction $R^{\rho}_{\phantom{\rho}\mu\rho\nu}\!=\!R_{\mu\nu}$ with contraction given by $R_{\eta\nu}g^{\eta\nu}\!=\!R$ as above: we then have that
\begin{eqnarray}
\nonumber
&G^{\rho}_{\phantom{\rho}\xi\mu\nu}=R^{\rho}_{\phantom{\rho}\xi\mu\nu}
+\frac{1}{2}(\nabla_{\mu}Q^{\rho}_{\phantom{\rho}\xi\nu}
\!-\!\nabla_{\nu}Q^{\rho}_{\phantom{\rho}\xi\mu})+\\
&+\frac{1}{4}(Q^{\rho}_{\phantom{\rho}\sigma\mu}Q^{\sigma}_{\phantom{\sigma}\xi\nu}
\!-\!Q^{\rho}_{\phantom{\rho}\sigma\nu}Q^{\sigma}_{\phantom{\sigma}\xi\mu})
\end{eqnarray}
is their most general decomposition. In the equivalent Lorentz formalism the curvature is written according to
\begin{eqnarray}
&G^{i}_{\phantom{i}j\mu\nu}
=\partial_{\mu}\omega^{i}_{j\nu}-\partial_{\nu}\omega^{i}_{j\mu}
+\omega^{i}_{p\mu}\omega^{p}_{j\nu}-\omega^{i}_{p\nu}\omega^{p}_{j\mu}
\end{eqnarray}
again antisymmetric in both the coordinate and the Lorentz indices: we have that we may write
\begin{eqnarray}
&G^{i}_{\phantom{i}j\mu\nu}\!=\!G^{\rho}_{\phantom{\rho}\sigma\mu\nu} e^{\sigma}_{j}e^{i}_{\rho}
\end{eqnarray}
as it should be for consistency. The gauge connection has an analogous curvature given by the similar expression
\begin{eqnarray}
&F_{\alpha\beta}=\partial_{\alpha}A_{\beta}-\partial_{\beta}A_{\alpha}
\end{eqnarray}
which is antisymmetric in its indices. Some additional identities for these curvatures are given by the following
\begin{eqnarray}
\nonumber
&D_{\mu}G^{\nu}_{\phantom{\nu}\iota \kappa \rho}
\!+\!D_{\kappa}G^{\nu}_{\phantom{\nu}\iota \rho \mu}
\!+\!D_{\rho}G^{\nu}_{\phantom{\nu}\iota \mu \kappa}+\\
&+G^{\nu}_{\phantom{\nu}\iota \beta \mu}Q^{\beta}_{\phantom{\beta}\rho \kappa}
\!+\!G^{\nu}_{\phantom{\nu}\iota \beta \kappa}Q^{\beta}_{\phantom{\beta}\mu \rho}
\!+\!G^{\nu}_{\phantom{\nu}\iota \beta \rho}Q^{\beta}_{\phantom{\beta}\kappa \mu}\equiv0
\label{cur}
\end{eqnarray}
again as it is easy to check directly: in Lorentz formalism this identity remains unchanged as expected; also the gauge curvature tensor verifies the Cauchy identities
\begin{eqnarray}
&\partial_{\nu}F_{\alpha\sigma}\!+\!\partial_{\sigma}F_{\nu\alpha}
\!+\!\partial_{\alpha}F_{\sigma\nu}=0
\end{eqnarray}
where despite the fact that the derivatives are ordinary partial derivatives such an expression is covariant indeed, as it is rather well known. The importance of the curvature tensor comes from the fact that with it, it becomes possible to express the commutator of covariant derivatives as $[D_{\mu},D_{\nu}]V^{\alpha}\!\!=\! Q^{\rho}_{\phantom{\rho}\mu\nu}D_{\rho}V^{\alpha}\!+\!
G^{\alpha}_{\phantom{\alpha}\rho\mu\nu}V^{\rho}$ in the case of vectors, and similarly with one curvature term for each tensorial index in the case of generic tensors of any order.

In terms of these geometrical quantities it is possible to build the spinorial equivalent of the curvature tensor
\begin{eqnarray}
&\boldsymbol{G}_{\mu\nu}
\!=\!\partial_{\mu}\boldsymbol{\Omega}_{\nu}\!-\!\partial_{\nu}\boldsymbol{\Omega}_{\mu}
\!+\!\boldsymbol{\Omega}_{\mu}\boldsymbol{\Omega}_{\nu}
\!-\!\boldsymbol{\Omega}_{\nu}\boldsymbol{\Omega}_{\mu}
\end{eqnarray}
antisymmetric in its two indices: it can be written as
\begin{eqnarray}
&\boldsymbol{G}_{\mu\nu}=\frac{1}{2}G^{ij}_{\phantom{ij}\mu\nu}\boldsymbol{\sigma}_{ij}
\!+\!iqF_{\mu\nu}\mathbb{I}
\end{eqnarray} 
in its most general decomposition. In terms of this curvature the commutator of spinorial covariant derivatives has the form $[\boldsymbol{D}_{\mu},\boldsymbol{D}_{\nu}]\psi\!=\! Q^{\rho}_{\phantom{\rho}\mu\nu} \boldsymbol{D}_{\rho}\psi\!+\!\boldsymbol{G}_{\mu\nu}\psi$ clearly showing that the gravitational and the electrodynamic curvatures may be formally united together within the most general curvature spinorial tensor that can be defined.
%%%%%%%%%%%%%%%%%%%%%%%%%%%%%%%%%%%%%%%%%%%%%%%%%%%%%%%%%%%%%%%%%%%%%%%%%%%%%%%%%%%%%%%%%%%%%%%%%%%
\subsection{Dynamical Equations}
Having defined the kinematic quantities, we will have these kinematic quantities coupled together by constructing the dynamical field equations we will employ next.

A first way we have to construct the most general system of field equations is to start from geometrical identities, postulating the geometric field equations that will convert these geometric identities into conservation laws, then postulating matter field equations that ensure those conservation laws be verified, and in this paper all these field equations will be taken at the least-order derivative that is possible: this geometric construction at the least-order derivative is easy, and so starting from the Jacobi-Bianchi geometrical identities for the torsion tensor and the curvature tensor (\ref{tor}-\ref{cur}) in their fully contracted form and geometrical identities obtained from the commutator of covariant derivatives applied to the case of the gauge strength $F_{\alpha\sigma}$ in their fully contracted form, it is possible to see that the least-order field equations, describing the coupling at the least-order derivative possible, are given for the coupling between the completely antisymmetric torsion and spin in a purely algebraic form as
\begin{eqnarray}
&Q^{\rho\mu\nu}=-kS^{\rho\mu\nu}
\end{eqnarray}
together with the field equations describing the coupling between the non-symmetric curvature and energy as
\begin{eqnarray}
\nonumber
&\left(\frac{1-k}{2k}\right)(D_{\mu}Q^{\mu\rho\alpha}
\!-\!\frac{1}{2}Q^{\theta\sigma\rho}Q_{\theta\sigma}^{\phantom{\theta\sigma}\alpha}
\!+\!\frac{1}{4}Q^{\theta\sigma\pi}Q_{\theta\sigma\pi}g^{\rho\alpha})+\\
\nonumber
&+(G^{\rho\alpha}\!-\!\frac{1}{2}Gg^{\rho\alpha}\!-\!\Lambda g^{\rho\alpha})-\\
&-\frac{1}{2}(\frac{1}{4}g^{\rho\alpha}F^{2}
\!-\!F^{\rho\theta}F^{\alpha}_{\phantom{\alpha}\theta})
=\frac{1}{2}T^{\rho\alpha}
\end{eqnarray}
and the field equations describing the coupling between the derivative of the gauge strength and the current
\begin{eqnarray}
&\frac{1}{2}F_{\mu\nu}Q^{\rho\mu\nu}\!+\!D_{\sigma}F^{\sigma\rho}=J^{\rho}
\end{eqnarray}
and these convert the above identities into conservation laws for the completely antisymmetric spin tensor as
\begin{eqnarray}
&D_{\rho}S^{\rho\mu\nu}\!+\!\frac{1}{2}(T^{\mu\nu}\!-\!T^{\nu\mu})=0
\end{eqnarray}
together with the conservation law for the non-symmetric energy tensor given according to the following form
\begin{eqnarray}
&D_{\mu}T^{\mu\rho}\!-\!T_{\mu\sigma}Q^{\sigma\mu\rho}\!+\!S_{\beta\mu\sigma}G^{\sigma\mu\beta\rho}
\!+\!J_{\beta}F^{\beta\rho}=0
\end{eqnarray}
and the conservation law for the current vector
\begin{eqnarray}
D_{\rho}J^{\rho}=0
\end{eqnarray}
which will have to be verified once the matter field equations are satisfied: these are verified by the conserved quantities given by the completely antisymmetric spin
\begin{eqnarray}
&S^{\rho\mu\nu}=a\frac{i}{4}\overline{\psi}\{\boldsymbol{\gamma}^{\rho}\!,\!\boldsymbol{\sigma}^{\mu\nu}\}\psi
\end{eqnarray}
with non-symmetric energy in the form
\begin{eqnarray}
\nonumber
&T^{\rho\alpha}=\frac{i}{2}(\overline{\psi}\boldsymbol{\gamma}^{\rho}\!\boldsymbol{D}^{\alpha}\psi
\!-\!\boldsymbol{D}^{\alpha}\overline{\psi}\!\boldsymbol{\gamma}^{\rho}\psi)-\\
\nonumber
&-(a\!-\!1)D_{\mu}(\frac{i}{4}\overline{\psi}\{\boldsymbol{\gamma}^{\mu}
\!,\!\boldsymbol{\sigma}^{\rho\alpha}\}\psi)+\\
\nonumber
&+(a\!-\!1)\frac{i}{4}\overline{\psi}\{\boldsymbol{\gamma}^{\rho}
\!,\!\boldsymbol{\sigma}_{\mu\nu}\}\psi Q^{\alpha\mu\nu}-\\
&-\frac{1}{2}(a\!-\!1)\frac{i}{4}\overline{\psi}\{\boldsymbol{\gamma}^{\alpha}
\!,\!\boldsymbol{\sigma}_{\mu\nu}\}\psi Q^{\rho\mu\nu}
\end{eqnarray}
and the current in terms of the expression
\begin{eqnarray}
&J^{\rho}=q\overline{\psi}\boldsymbol{\gamma}^{\rho}\psi
\end{eqnarray}
whenever the set of least-order field equations for the fermion fields given according to the expressions
\begin{eqnarray}
&i\boldsymbol{\gamma}^{\mu}\!\boldsymbol{D}_{\mu}\psi
\!+\!\frac{i}{8}(a\!-\!1)Q_{\rho\mu\nu}\boldsymbol{\gamma}^{\rho}\boldsymbol{\gamma}^{\mu}
\boldsymbol{\gamma}^{\nu}\psi\!-\!m\psi\!=\!0
\end{eqnarray}
are satisfied as conditions on the fermion matter fields.

This geometric construction follows the spirit of deriving from geometry all physical equations, but it is also possible to write the most general least-order derivative Lagrangian, and then vary it to obtain the least-order derivative field equations: so far as we know, the most general torsional-gravitational dynamics for matter fields is given by the Lagrangian obtained by combining the results of \cite{Hojman:1980kv} and \cite{Alexandrov:2008iy}, and for axial vector torsion it reduces to the one in \cite{Fabbri:2012yg,Fabbri:2013gza}; in addition, we will also include all possible interacting terms between all fields involved, which for an axial vector torsion means that there is a single supplementary term given as a product between torsion and the spinor field \cite{Fabbri:2014naa}, and the Lagrangian
\begin{eqnarray}
\nonumber
&L\!=\!(\frac{k-1}{4k})Q_{\alpha\nu\sigma}Q^{\alpha\nu\sigma}\!+\!G\!+\!2\Lambda
\!+\!\frac{1}{4}F^{\alpha\nu}F_{\alpha\nu}-\\
\nonumber
&-\frac{i}{2}(\overline{\psi}\boldsymbol{\gamma}^{\mu}\!\boldsymbol{D}_{\mu}\psi
\!-\!\boldsymbol{D}_{\mu}\overline{\psi}\!\boldsymbol{\gamma}^{\mu}\psi)-\\
&-\frac{1}{8}(a\!-\!1)i\overline{\psi}\boldsymbol{\gamma}^{\nu} \boldsymbol{\gamma}^{\sigma}\boldsymbol{\gamma}^{\pi}\psi Q_{\nu\sigma\pi}
\!+\!m\overline{\psi}\psi
\end{eqnarray}
is the most general least-order derivative Lagrangian up to dimension four that can be written in our case.

Given this dynamical action in its most extensively coupled form, its variation yields the system of dynamical field equations for the completely antisymmetric torsion-spin coupling as given according to the expression
\begin{eqnarray}
&Q^{\rho\mu\nu}\!=\!\frac{1}{4}ak\overline{\psi}\boldsymbol{\gamma}_{\sigma}\boldsymbol{\pi}\psi
\varepsilon^{\sigma\rho\mu\nu}
\end{eqnarray}
which come together with the field equations for the non-symmetric curvature-energy coupling according to
\begin{eqnarray}
\nonumber
&\left(\frac{1-k}{2k}\right)(D_{\mu}Q^{\mu\rho\alpha}
\!-\!\frac{1}{2}Q^{\theta\sigma\rho}Q_{\theta\sigma}^{\phantom{\theta\sigma}\alpha}
\!+\!\frac{1}{4}Q^{\theta\sigma\pi}Q_{\theta\sigma\pi}g^{\rho\alpha})+\\
\nonumber
&+(G^{\rho\alpha}\!-\!\frac{1}{2}Gg^{\rho\alpha}\!-\!\Lambda g^{\rho\alpha})-\\
\nonumber
&-\frac{1}{2}(\frac{1}{4}g^{\rho\alpha}F^{2}
\!-\!F^{\rho\theta}F^{\alpha}_{\phantom{\alpha}\theta})=\\
\nonumber
&=\frac{i}{4}(\overline{\psi}\boldsymbol{\gamma}^{\rho}\!\boldsymbol{D}^{\alpha}\psi
\!-\!\boldsymbol{D}^{\alpha}\overline{\psi}\!\boldsymbol{\gamma}^{\rho}\psi)-\\
\nonumber
&-\frac{1}{2}(a\!-\!1)D_{\mu}(\frac{i}{4}\overline{\psi}\{\boldsymbol{\gamma}^{\mu}
\!,\!\boldsymbol{\sigma}^{\rho\alpha}\}\psi)+\\
\nonumber
&+\frac{1}{2}(a\!-\!1)\frac{i}{4}\overline{\psi}\{\boldsymbol{\gamma}^{\rho}
\!,\!\boldsymbol{\sigma}_{\mu\nu}\}\psi Q^{\alpha\mu\nu}-\\
&-\frac{1}{4}(a\!-\!1)\frac{i}{4}\overline{\psi}\{\boldsymbol{\gamma}^{\alpha}
\!,\!\boldsymbol{\sigma}_{\mu\nu}\}\psi Q^{\rho\mu\nu}
\end{eqnarray}
and the field equations for the gauge-current coupling
\begin{eqnarray}
&\frac{1}{2}F_{\mu\nu}Q^{\rho\mu\nu}\!+\!D_{\sigma}F^{\sigma\rho}
\!=\!q\overline{\psi}\boldsymbol{\gamma}^{\rho}\psi
\end{eqnarray}
complemented by the fermionic field equations
\begin{eqnarray}
&i\boldsymbol{\gamma}^{\mu}\!\boldsymbol{D}_{\mu}\psi
\!+\!\frac{1}{8}(a\!-\!1)\varepsilon^{\rho\mu\nu\sigma}
Q_{\rho\mu\nu}\boldsymbol{\gamma}_{\sigma}\boldsymbol{\pi}\psi\!-\!m\psi\!=\!0
\end{eqnarray}
in which the torsional constant $k$ is not the gravitational constant, which has been set to the unity, and the constant $a$ measures the strength with which torsion couples to fermions. This system of field equations coincides with the one we have found above with a geometrical spirit.

Because torsion is a tensor, all curvatures and derivatives are decomposed in terms of torsionless curvatures and derivatives plus torsional contributions, and since the torsion-spin field equations are algebraic, the torsional contributions can be substituted in terms of the spin of fermionic fields: the conservation law for the spin accounts for the antisymmetric part of the non-symmetric curvature-energy field equations, and so the conservation law for the symmetric torsionless energy is given by
\begin{eqnarray}
&\nabla_{\mu}E^{\mu\rho}\!+\!J_{\beta}F^{\beta\rho}=0
\end{eqnarray}
while the conservation law for the current vector is
\begin{eqnarray}
\nabla_{\rho}J^{\rho}=0
\end{eqnarray}
verified by the matter field equations: the conserved quantities are given by the symmetric torsionless energy
\begin{eqnarray}
\nonumber
&E^{\rho\alpha}\!=\!
\frac{i}{4}(\overline{\psi}\boldsymbol{\gamma}^{\rho}\boldsymbol{\nabla}^{\alpha}\psi
\!-\!\boldsymbol{\nabla}^{\alpha}\overline{\psi}\boldsymbol{\gamma}^{\rho}\psi+\\
\nonumber
&+\overline{\psi}\boldsymbol{\gamma}^{\alpha}\boldsymbol{\nabla}^{\rho}\psi
\!-\!\boldsymbol{\nabla}^{\rho}\overline{\psi}\boldsymbol{\gamma}^{\alpha}\psi)+\\
&+\frac{1}{2}Y\overline{\psi}\boldsymbol{\gamma}_{\mu}\boldsymbol{\pi}\psi
\overline{\psi}\boldsymbol{\gamma}^{\mu}\boldsymbol{\pi}\psi g^{\alpha\rho}
\end{eqnarray}
and the current in terms of the expression
\begin{eqnarray}
&J^{\rho}=q\overline{\psi}\boldsymbol{\gamma}^{\rho}\psi
\end{eqnarray}
whenever the field equations for the fermion fields
\begin{eqnarray}
&i\boldsymbol{\gamma}^{\mu}\boldsymbol{\nabla}_{\mu}\psi
\!+\!Y\overline{\psi}\boldsymbol{\gamma}_{\rho}\boldsymbol{\pi}\psi
\boldsymbol{\gamma}^{\rho}\boldsymbol{\pi}\psi\!-\!m\psi\!=\!0
\end{eqnarray}
are satisfied as conditions on the fermion matter fields.

Also the action can be written in the form in which everything is decomposed in terms of the torsionless counterparts plus torsional contributions substituted in terms of the spin of the fermionic fields yielding
\begin{eqnarray}
\nonumber
&L=R\!+\!2\Lambda\!+\!\frac{1}{4}F^{\mu\nu}F_{\mu\nu}-\\
\nonumber
&-\frac{i}{2}(\overline{\psi}\boldsymbol{\gamma}^{\mu}\boldsymbol{\nabla}_{\mu}\psi
\!-\!\boldsymbol{\nabla}_{\mu}\overline{\psi}\boldsymbol{\gamma}^{\mu}\psi)-\\
&-\frac{1}{2}Y\overline{\psi}\boldsymbol{\gamma}_{\mu}\boldsymbol{\pi}\psi
\overline{\psi}\boldsymbol{\gamma}^{\mu}\boldsymbol{\pi}\psi
\!+\!m\overline{\psi}\psi
\end{eqnarray}
which will eventually yield the system of the field equations already in the form in which torsion has been replaced with spin-spin contact fermionic interactions.

Or alternatively, these field equations can be obtained from the previous field equations after decomposing all curvatures and derivatives into the corresponding torsionless curvatures and derivatives plus torsional contributions written as the spin of fermions, yielding the symmetric curvature-energy coupling field equations as
\begin{eqnarray}
\nonumber
&(R^{\rho\alpha}\!-\!\frac{1}{2}Rg^{\rho\alpha}\!-\!\Lambda g^{\rho\alpha})
\!-\!\frac{1}{2}(\frac{1}{4}g^{\rho\alpha}F^{2}
\!-\!F^{\rho\theta}F^{\alpha}_{\phantom{\alpha}\theta})=\\
\nonumber
&=\frac{i}{8}(\overline{\psi}\boldsymbol{\gamma}^{\rho}\boldsymbol{\nabla}^{\alpha}\psi
\!-\!\boldsymbol{\nabla}^{\alpha}\overline{\psi}\boldsymbol{\gamma}^{\rho}\psi+\\
\nonumber
&+\overline{\psi}\boldsymbol{\gamma}^{\alpha}\boldsymbol{\nabla}^{\rho}\psi
\!-\!\boldsymbol{\nabla}^{\rho}\overline{\psi}\boldsymbol{\gamma}^{\alpha}\psi)+\\
&+\frac{1}{4}Y\overline{\psi}\boldsymbol{\gamma}_{\mu}\boldsymbol{\pi}\psi
\overline{\psi}\boldsymbol{\gamma}^{\mu}\boldsymbol{\pi}\psi g^{\alpha\rho}
\label{gravitational}
\end{eqnarray}
and with the gauge-current coupling field equations
\begin{eqnarray}
&\nabla_{\sigma}F^{\sigma\rho}\!=\!q\overline{\psi}\boldsymbol{\gamma}^{\rho}\psi
\label{electrodynamical}
\end{eqnarray}
together with the fermionic field equations
\begin{eqnarray}
&i\boldsymbol{\gamma}^{\mu}\boldsymbol{\nabla}_{\mu}\psi
\!-\!Y\overline{\psi}\boldsymbol{\gamma}_{\rho}\psi
\boldsymbol{\gamma}^{\rho}\psi\!-\!m\psi\!=\!0
\label{fermionical}
\end{eqnarray}
showing that the system of field equations has reduced to the one we would have had without torsion but supplemented with non-linear potentials of a specific structure and with self-coupling constant $3ka^{2}\!=\!16Y$ showing that the two constants related to torsion are reexpressed as a single constant describing the self-coupling of the matter.

This theory is the only torsional-completion of gravity with matter fields that is compatible with all experiments as it has been discussed in \cite{Fabbri:2014vda} and references therein.

The non-linearity terms have effects that are interestingly similar to those due to quantum corrections.
%%%%%%%%%%%%%%%%%%%%%%%%%%%%%%%%%%%%%%%%%%%%%%%%%%%%%%%%%%%%%%%%%%%%%%%%%%%%%%%%%%%%%%%%%%%%%%%%%%%
%%%%%%%%%%%%%%%%%%%%%%%%%%%%%%%%%%%%%%%%%%%%%%%%%%%%%%%%%%%%%%%%%%%%%%%%%%%%%%%%%%%%%%%%%%%%%%%%%%%
\section{Non-Linearity Terms}
%%%%%%%%%%%%%%%%%%%%%%%%%%%%%%%%%%%%%%%%%%%%%%%%%%%%%%%%%%%%%%%%%%%%%%%%%%%%%%%%%%%%%%%%%%%%%%%%%%%
\subsection{Quantum-like Corrections}
As discussed in the introduction, quantum field theory has been able to produce predictions with impressive precision, but regardless of how many and however precise these predictions are, they cannot be taken as reasons to accept the theory beyond any doubt, while on the other hand this theory has also produced some astonishingly wrong predictions which must be considered as evidence of incorrectness; also to remark is the fact that, because quantum field theory is based on a larger set of axioms compared to torsion-gravitational matter field theory, then it follows that the former is theoretically weaker than the latter: as already stated, our position will be that of renouncing to have any protocol of quantization in the first place. Despite renouncing to quantization may be considered as extreme, nevertheless we already discussed that it is equivalent to implementing the normal-ordering procedure: in quantum field theory this is done when convenient, and in the following we will assume such prescription to be done not only when convenient but in general. Renouncing to field quantization by not assuming any commutation relationships between operatorial re-definitions of fields means that there is no need to have fields re-defined in terms of operators of creation and annihilation. Therefore, if we have no quantization to produce the observed effects then such effects must be recovered in terms of something else like non-linearity terms in the matter field equations.

A very first point we have to consider is that assuming no field quantization means that all problems coming as consequence of it will drop: in particular, the problem of the cosmological constant contribution due to the zero-point energy would not even arise; in addition, we take here the opportunity to recall that also contributions due to spontaneous-breaking of a given symmetry can be circumvented in a model in which it is not with gauge interactions that the phenomenology of the weak forces of lepton fields is obtained. Basically, this idea of describing weak-like forces for lepton fields merely consists in applying the Nambu-Jona--Lasinio analysis to the case of two leptons so to have the weak mediators as condensed states of lepton fields \cite{n-j--l/1,n-j--l/2, g-n,s-g,Fabbri:2010ux}, and in a similar manner such an idea can also be used to describe oscillations among neutrinos even if they are massless and single-handed \cite{a-b,w,ds-g, Fabbri:2010hz}; condensed states of neutral massive particles can be used at galactic scales in order to fit the behaviour of the rotation curves \cite{Tilquin:2011bu, Silverman:2002qx, Fabbri:2012zc}. And non-linear contributions solve the problem of gravitationally-induced singularity formation \cite{k,Fabbri:2011mg}. It is in fact quite interesting that some of the open problems of cosmology can be addressed in terms of non-linearity potentials while disregarding fermionic-field quantization \cite{Magueijo:2012ug}.

However, the problem of having no field quantization, and the consequent lack of zero-point energy, means that one needs to find a way to recover the effects due to the vacuum fluctuations; in particular, there must be a way to recover the Casimir effect without employing the zero-point energy: this can be done in terms of the effective interactions with the electrodynamic sources constituted by the two conducting plates. In fact, Casimir himself and his collaborators originally derived their results as the effects of retarded van der Waals forces acting between the metallic plates; although later they found that the same results could be derived more simply in terms of a heuristic argument based on vacuum energy, Casimir forces can be obtained as result of effective interactions between electrodynamic currents \cite{j}. Furthermore, if in this paper Jaffe describes the Casimir force without zero-point lowest-order quantum correction but only in terms of higher-order quantum corrections involving radiative processes connected with external legs, in a more drastic possibility Schwinger describes the Casimir force without any radiative process at all, but only in terms of classical fields in interaction with given external sources \cite{Schwinger:1989ix}.

The assumption of obtaining quantum effects in terms of classical fields interacting with external sources may sound extreme, but nevertheless this is precisely what Schwinger's theory does \cite{s,Rugh:2000ji}. On the other hand, for fermions having torsionally-induced non-linear potentials, it has been discussed, in \cite{Fabbri:2010rw}, that the Pauli principle may be entailed, in  \cite{Fabbri:2012ag}, that the positivity of the energy is ensured, in \cite{Fabbri:2013isa}, that there is no time-reversal symmetry, and finally that the $4$-Fermi spin-spin forces mimic the effects of quantum corrections. Here, we would like to recover these results in a manner that is closer to the work of Schwinger. Namely, showing that even for extended fields in self-interaction through their own electrodynamic field we may get such quantum corrections.

To this purpose, we introduce the following definitions of all possible bi-linear spinorial fields
\begin{eqnarray}
&S_{\mu\nu}\!=\!i\overline{\psi}\boldsymbol{\sigma}_{\mu\nu}\psi\\
&\Sigma_{\mu\nu}\!=\!\overline{\psi}\boldsymbol{\sigma}_{\mu\nu}\boldsymbol{\pi}\psi\\
&P^{\nu}\!=\!\overline{\psi}\boldsymbol{\gamma}^{\nu}\psi\\
&V^{\nu}\!=\!\overline{\psi}\boldsymbol{\gamma}^{\nu}\boldsymbol{\pi}\psi\\
&\Phi\!=\!\overline{\psi}\psi\\
&\Theta\!=\!i\overline{\psi}\boldsymbol{\pi}\psi
\end{eqnarray}
such that we have the Fierz identities given in the form
\begin{eqnarray}
\nonumber
&4S_{\mu\nu}S^{\alpha\nu}\!-\!\Phi^{2}\delta_{\mu}^{\alpha}
\!=\!4\Sigma_{\mu\nu}\Sigma^{\alpha\nu}\!-\!\Theta^{2}\delta_{\mu}^{\alpha}=\\
&=V_{\mu}V^{\alpha}\!\!-\!P_{\mu}P^{\alpha}\\
\nonumber
&\!P_{\mu}P^{\mu}\!=\!-V_{\mu}V^{\mu}=\\
&=\Phi^{2}\!+\!\Theta^{2}\\
&\!4\Sigma_{\mu\nu}S^{\alpha\nu}\!=\!-\Theta\Phi\delta_{\mu}^{\alpha}\\
&P_{\mu}V^{\mu}\!=\!0
\end{eqnarray}
together with
\begin{eqnarray}
&2S_{\alpha\mu}P^{\alpha}=\Theta V_{\mu}\\
&2\Sigma_{\alpha\mu}P^{\alpha}=\Phi V_{\mu}\\
&2S_{\alpha\mu}V^{\alpha}=\Theta P_{\mu}\\
&2\Sigma_{\alpha\mu}V^{\alpha}=\Phi P_{\mu}
\end{eqnarray}
and also
\begin{eqnarray}
&\Phi S_{\mu\nu}\!-\!\Theta\Sigma_{\mu\nu}
\!=\!\frac{1}{2}P^{\alpha}V^{\sigma}\varepsilon_{\alpha\sigma\mu\nu}
\end{eqnarray}
as geometrical identities; the six definitions of the bi-linear spinorial fields have been given in such a way that the subsequent Fierz identities could be written in a symmetric way, but identity $S_{\mu\nu}\!=\!\frac{1}{2}\varepsilon_{\mu\nu\alpha\beta}\Sigma^{\alpha\beta}$ tells that these two quantities are in fact dependent, and therefore there are in total only $6\!+\!4\!+\!4\!+\!
1\!+\!1\!=\!16$ independent components as expected for four-dimensional spinorial fields.

For the aim we have in mind, in the following we will try to see what are the properties of the system by studying one trial solution: a simple guess is given by
\begin{eqnarray}
&i\boldsymbol{\nabla}_{\mu}\psi\!-\!(Y\!+\!1\!-\!b)P_{\mu}\psi
\!-\!b\frac{m}{4}\boldsymbol{\gamma}_{\mu}\psi=0
\label{mattersolution}
\end{eqnarray}
constrained by $P_{\mu}\boldsymbol{\gamma}^{\mu}\psi\!=\!m\psi$ with $P^{2}\!=\!m^{2}$ and $b$ a generic constant still unspecified: as it is usual for plane-wave solutions, the constraint $P_{\mu}\boldsymbol{\gamma}^{\mu}\psi\!=\!m\psi$ implies other constraints such as $\Phi\!=\!m$ and $\Theta\!=\!0$ with the consequence that in the zero-mass configuration both bi-linear scalars vanish and therefore the spinor can only have a single-handed chirality; once this form is plugged, and the ensuing constraints are considered, within the matter field equations, it is possible to check that this is indeed an implicit exact solution of the matter field equations (\ref{fermionical}).

With the solution (\ref{mattersolution}) we also have that
\begin{eqnarray}
&A_{\mu}\!=\!-\frac{2q}{3m^{2}b^{2}}P_{\mu}
\label{gaugesolution}
\end{eqnarray}
solves the electrodynamic field equations: to see that we write the solution (\ref{mattersolution}) in the normal form as
\begin{eqnarray}
&\boldsymbol{\nabla}_{\mu}\psi\!=\!-i(Y\!+\!1\!-\!b)P_{\mu}\psi
\!-\!ib\frac{m}{4}\boldsymbol{\gamma}_{\mu}\psi\\
&\boldsymbol{\nabla}_{\mu}\overline{\psi}\!=\!i(Y\!+\!1\!-\!b)P_{\mu}\overline{\psi}
\!+\!ib\frac{m}{4}\overline{\psi}\boldsymbol{\gamma}_{\mu}
\end{eqnarray}
from which it is possible to see first that
\begin{eqnarray}
\nonumber
&\boldsymbol{\nabla}_{\mu}(\overline{\psi}\boldsymbol{\gamma}_{\nu}\psi)=\\
\nonumber
&=\boldsymbol{\nabla}_{\mu}\overline{\psi}\boldsymbol{\gamma}_{\nu}\psi
\!+\!\overline{\psi}\boldsymbol{\gamma}_{\nu}\boldsymbol{\nabla}_{\mu}\psi=\\
\nonumber
&=i(Y\!+\!1\!-\!b)P_{\mu}\overline{\psi}\boldsymbol{\gamma}_{\nu}\psi
\!+\!ib\frac{m}{4}\overline{\psi}\boldsymbol{\gamma}_{\mu}\boldsymbol{\gamma}_{\nu}\psi-\\
\nonumber
&-i(Y\!+\!1\!-\!b)P_{\mu}\overline{\psi}\boldsymbol{\gamma}_{\nu}\psi
\!-\!ib\frac{m}{4}\overline{\psi}\boldsymbol{\gamma}_{\nu}\boldsymbol{\gamma}_{\mu}\psi=\\
\nonumber
&=ib\frac{m}{4}\overline{\psi}[\boldsymbol{\gamma}_{\mu},\boldsymbol{\gamma}_{\nu}]\psi=\\
&=ibm\overline{\psi}\boldsymbol{\sigma}_{\mu\nu}\psi
\label{aux}
\end{eqnarray}
and then also that
\begin{eqnarray}
\nonumber
&\boldsymbol{\nabla}_{\mu}
[\boldsymbol{\nabla}^{[\mu}(\overline{\psi}\boldsymbol{\gamma}^{\nu]}\psi)]=\\
\nonumber
&=2bm\boldsymbol{\nabla}_{\mu}(i\overline{\psi}\boldsymbol{\sigma}^{\mu\nu}\psi)=\\
\nonumber
&=2ibm(\boldsymbol{\nabla}_{\mu}\overline{\psi}\boldsymbol{\sigma}^{\mu\nu}\psi
\!+\!\overline{\psi}\boldsymbol{\sigma}^{\mu\nu}\boldsymbol{\nabla}_{\mu}\psi)=\\
\nonumber
&=2ibm[i(Y\!+\!1\!-\!b)P_{\mu}\overline{\psi}\boldsymbol{\sigma}^{\mu\nu}\psi
\!+\!ib\frac{m}{4}\overline{\psi}\boldsymbol{\gamma}_{\mu}\boldsymbol{\sigma}^{\mu\nu}\psi-\\
\nonumber
&-i(Y\!+\!1\!-\!b)P_{\mu}\overline{\psi}\boldsymbol{\sigma}^{\mu\nu}\psi
\!-\!ib\frac{m}{4}\overline{\psi}\boldsymbol{\sigma}^{\mu\nu}\boldsymbol{\gamma}_{\mu}\psi]=\\
\nonumber
&=-\frac{1}{2}b^{2}m^{2}
\overline{\psi}[\boldsymbol{\gamma}_{\mu},\boldsymbol{\sigma}^{\mu\nu}]\psi=\\
\nonumber
&=-\frac{1}{2}b^{2}m^{2}\overline{\psi}(4\boldsymbol{\gamma}^{\nu}
\!-\!g^{\mu\nu}\boldsymbol{\gamma}_{\mu})\psi=\\
&=-\frac{3}{2}b^{2}m^{2}\overline{\psi}\boldsymbol{\gamma}^{\nu}\psi
\end{eqnarray}
showing that (\ref{gaugesolution}) is solution because it converts the last relationship into the electrodynamic field equations (\ref{electrodynamical}).

Analogously, (\ref{mattersolution}) and (\ref{gaugesolution}) can be employed to see that
\begin{eqnarray}
\nonumber
&R_{\rho\alpha}\!+\!\Lambda g_{\rho\alpha}
\!=\!-\frac{2q^{2}}{9m^{2}b^{2}}V_{\rho}V_{\alpha}+\\
\nonumber
&+(\frac{2q^{2}}{9m^{2}b^{2}}\!-\!\frac{b}{2}\!+\!\frac{Y}{2}\!+\!\frac{1}{2})P_{\rho}P_{\alpha}-\\
&-(\frac{q^{2}}{9m^{2}b^{2}}\!-\!\frac{b}{8}\!+\!\frac{1}{4})m^{2}g_{\rho\alpha}
\end{eqnarray}
are in fact the gravitational field equations (\ref{gravitational}).

In the system of coordinates in which locally the gravitational field is negligible, the derivative of the matter field only contains the contributions of the gauge field, and by means of (\ref{mattersolution}) and (\ref{gaugesolution}) it is possible to see that the interaction terms can be converted into
\begin{eqnarray}
\nonumber
&-(qA_{\mu}\!+\!Y\overline{\psi}\boldsymbol{\gamma}_{\mu}\psi)\boldsymbol{\gamma}^{\mu}\psi=\\
\nonumber
&=(\frac{2q^{2}}{3b^{2}m^{2}}\!-\!Y)
\overline{\psi}\boldsymbol{\gamma}_{\mu}\psi\boldsymbol{\gamma}^{\mu}\psi=\\
\nonumber
&=2(\frac{2q^{2}}{3b^{2}m^{2}}\!-\!Y)i\overline{\psi}\boldsymbol{\sigma}_{\mu\nu}\psi i\boldsymbol{\sigma}^{\mu\nu}\psi=\\
&=-\frac{i}{2}(\frac{2}{b}\!-\!3Yb\frac{m^{2}}{q^{2}})
\frac{q}{m}F_{\mu\nu}\boldsymbol{\sigma}^{\mu\nu}\psi
\end{eqnarray}
where $\overline{\psi}\boldsymbol{\gamma}_{\mu}\psi\boldsymbol{\gamma}^{\mu}\psi
\!\equiv\!2i\overline{\psi}\boldsymbol{\sigma}_{\mu\nu}\psi i\boldsymbol{\sigma}^{\mu\nu}\psi
\!+\!2i\overline{\psi}\boldsymbol{\pi}\psi i\boldsymbol{\pi}\psi$ was used and the constraint $i\overline{\psi}\boldsymbol{\pi}\psi\!=\!0$ taken into account throughout the entire calculation; having Fierz rearranged in this manner the interacting term we end up with 
\begin{eqnarray}
&i\boldsymbol{\gamma}^{\mu}\boldsymbol{\nabla}_{\mu}\psi
\!-\!\frac{i}{2}\!\left(\frac{2}{b}\!-\!3Yb\frac{m^{2}}{q^{2}}\right)\!
\frac{q}{m}F_{\mu\nu}\boldsymbol{\sigma}^{\mu\nu}\psi\!-\!m\psi\!=\!0
\end{eqnarray}
as the final form of the matter field equation.

Notice that to get this form we have also used relation
\begin{eqnarray}
&F_{\mu\nu}\!=\!-\frac{4q}{3mb}S_{\mu\nu}
\end{eqnarray}
obtained from (\ref{aux}) and representing the fact that the fermionic spin generates a magnetic field, as it is known.

We see that in the rearranged matter field equation
\begin{eqnarray}
&i\boldsymbol{\gamma}^{\mu}\boldsymbol{\nabla}_{\mu}\psi
\!-\!\frac{i}{2}\!\left(\frac{2}{b}\!-\!3Yb\frac{m^{2}}{q^{2}}\right)\!
\frac{q}{m}F_{\mu\nu}\boldsymbol{\sigma}^{\mu\nu}\psi\!-\!m\psi\!=\!0
\end{eqnarray}
there arise Pauli-types of corrections.

Deriving Pauli-type correction in terms of extended fermions interacting with their own electrodynamic field, beside the contributions due to the self-coupling mediated by torsion, is important because it means that these quantum-like effects may have an intrinsically electrodynamic essence, beyond the influence of torsion, compatibly with the computation of radiative processes for external fields, in non-trivial background, as it has been presented in works \cite{Buchbinder:1985ym,Buchbinder:1987eh} and \cite{Buchbinder:1989zz}. The fundamental idea is that self-interacting fields of matter may be characterized by effective dynamics that are analogous to those commonly ascribed to quantization prescriptions \cite{Buchbinder:1992rb}.

It is to be noticed that here there is no perturbative expansion and therefore we can be sure that the result is finite as it was also in \cite{Fabbri:2013isa}, while in the common approach and in Schwinger source theory approach the Pauli term is computed in perturbative expansions which are not known to converge: results can be made finite order by order, and for the orders that can be computed both the common approach and Schwinger's approach are capable of predicting the correct correction \cite{s}, while here the specific parameter relies upon the choice of a specific form of the solution, and therefore it cannot be calculated precisely because we decided to discuss the problem regardless the particular shape acquired by the extended matter field distribution. That we cannot gain any more information apart from the finiteness of the interaction is a consequence of our will to study the problem in general.

The fact of dealing with extended matter field distributions implies they will tend to spread over the entire spacetime then rising the question about how they can be localized within a compact region, and in general the answer can be given in two ways: one is to take advantage of the fact that the torsional coupling constant is not fixed so that we may assume it is negative and with an opportunely large value, resulting in an attraction strong enough to ensure localized stable solutions, as it is done for the case of solitons \cite{t}; the other is to focus on the universal attractiveness of the gravitational field.

In the following, we are going to maintain all potentials within the matter field equations: from the Maurer-Cartan equation $e_{a}^{\rho}e^{\mu}_{b}e^{\sigma}_{c}(\partial_{\rho}e_{\mu}^{c}
\!-\!\partial_{\mu}e_{\rho}^{c})\!\equiv\!-C_{ab}^{\phantom{ab}k}e^{\sigma}_{k}$ we define
\begin{eqnarray}
&C^{\alpha}\!=\!-\frac{1}{8}C_{\nu\sigma\pi}\varepsilon^{\nu\sigma\pi\alpha}
\!=\!\frac{1}{4}\partial_{\mu}e_{\rho}^{k}e_{\sigma}^{j}\eta_{jk}
\varepsilon^{\mu\rho\sigma\alpha}\\
&B_{k}\!=\!\frac{1}{2}\frac{1}{\sqrt{|g|}}\partial_{\mu}(\sqrt{|g|}e^{\mu}_{k})
\label{potentials}
\end{eqnarray}
where, under coordinate and conformal transformations, the first gravitational potential $C_{k}$ is covariant so that we have to think at $C_{k}$ as what encodes the distortion due to tidal effects, while $B_{k}$ encodes the compression due to gravitational effects, and $A_{\mu}$ is the electrodynamic potential as usual: after some Fierz rearrangement, the field equations (\ref{fermionical}) can be written in the following form
\begin{eqnarray}
\nonumber
&i\boldsymbol{\gamma}^{\mu}\partial_{\mu}\psi
\!+\![(Y\!-\!X)V_{\mu}\!+\!C_{\mu}]\boldsymbol{\gamma}^{\mu}\boldsymbol{\pi}\psi
\!+\!iB_{\mu}\boldsymbol{\gamma}^{\mu}\psi-\\
&-(XP_{\mu}\!+\!qA_{\mu})\boldsymbol{\gamma}^{\mu}\psi\!-\!m\psi\!=\!0
\label{explicit}
\end{eqnarray}
with non-linear terms; this non-linear Dirac equation has non-relativistic limit given by quintic Pauli-Schr\"{o}dinger equations, which are non-integrable. However, by employing special formal solutions, it may be possible to obtain additional information about some properties that, nevertheless, may be valid in general circumstances.

For example, it is well known that a formal solution of field equations (\ref{explicit}) may be obtained in terms of the factorization that is given according to the form
\begin{eqnarray}
\nonumber
&\psi\!=\![\cos{(\int[(Y\!\!-\!\!X)V_{\nu}\!+\!C_{\nu}]dx^{\nu})}+\\
\nonumber
&+i\boldsymbol{\pi}\sin{(\int[(Y\!\!-\!\!X)V_{\nu}\!+\!C_{\nu}]dx^{\nu})}]\cdot\\
\nonumber
&\cdot[\cos{(\int(XP_{\mu}\!+\!qA_{\mu})dx^{\mu})}-\\
\nonumber
&-i\sin{(\int(XP_{\mu}\!+\!qA_{\mu})dx^{\mu})}]\cdot\\
&\cdot\exp{(-\!\int\!B_{\alpha}dx^{\alpha})}\phi
\label{solution}
\end{eqnarray}
where 
\begin{eqnarray}
&P_{\mu}\!\equiv\!\overline{\psi}\boldsymbol{\gamma}_{\mu}\psi
\!=\!\overline{\phi}\boldsymbol{\gamma}_{\mu}\phi
\exp{(-2\int\!B_{\alpha}dx^{\alpha})}
\label{solvector}\\
&V_{\mu}\!\equiv\!\overline{\psi}\boldsymbol{\gamma}_{\mu}\boldsymbol{\pi}\psi
\!=\!\overline{\phi}\boldsymbol{\gamma}_{\mu}\boldsymbol{\pi}\phi
\exp{(-2\int\!B_{\alpha}dx^{\alpha})}
\label{solaxialvec}
\end{eqnarray}
and with the spinor $\phi$ solution of 
\begin{eqnarray}
\nonumber
&i\boldsymbol{\gamma}^{\mu}\partial_{\mu}\phi
\!-\!m[\cos{(2\int[(Y\!\!-\!\!X)V_{\nu}\!+\!C_{\nu}]dx^{\nu})}+\\
&+i\boldsymbol{\pi}\sin{(2\int[(Y\!\!-\!\!X)V_{\nu}\!+\!C_{\nu}]dx^{\nu})}]\phi\!=\!0
\label{ffe}
\end{eqnarray}
which can then be treated with the same methods we have exposed above: we notice that the effects of the vector spinor and the electrodynamic gauge fields are those of producing an oscillation among the real and complex parts of the spinorial components while the effects of the axial-vector bi-linear spinor and the tidal potential fields are those of producing an oscillation among real and complex components but differently for the left-handed and right-handed semi-spinorial chiral projections, and finally the effect of the gravitational potential is that of producing an exponential damping. So soon as the gravitational damping is strong enough to balance the spreading of the extended matter field distribution, confinement occurs.

As a consequence of the confinement, we study the conditions that define the shape of the region where the spinor is at its maximal value: in such a region of maximal value, we can consider it to be nearly constant, and because this localized distribution is massive then we may always boost in its rest frame where we have that the constraint $\Theta\!=\!0$ holds, so that the field equations become
\begin{eqnarray}
\nonumber
&[(Y\!-\!X)V_{\mu}\!+\!C_{\mu}]\boldsymbol{\gamma}^{\mu}\boldsymbol{\pi}\psi
\!+\!iB_{\mu}\boldsymbol{\gamma}^{\mu}\psi-\\
&-(XP_{\mu}\!+\!qA_{\mu})\boldsymbol{\gamma}^{\mu}\psi
\!-\!m\psi\!=\!0
\end{eqnarray}
showing that apart the trivial solution $\psi\!\equiv\!0$ there is also the solution for which the spinor is maximum; this is given when the spinor is not equal to zero but given as
\begin{eqnarray}
\nonumber
&\mathrm{det}|[(Y\!-\!X)V_{\mu}\!+\!C_{\mu}]\boldsymbol{\gamma}^{\mu}\boldsymbol{\pi}
\!+\!iB_{\mu}\boldsymbol{\gamma}^{\mu}-\\
&-(XP_{\mu}\!+\!qA_{\mu})\boldsymbol{\gamma}^{\mu}\!-\!m|\!=\!0
\end{eqnarray}
resulting into
\begin{eqnarray}
&2C_{\mu}\Sigma^{\mu\alpha}\!+\!2B_{\mu}S^{\mu\alpha}
\!+\!qA^{\alpha}\Phi\!+\!(Y\Phi\!+\!m)P^{\alpha}\!=\!0\\
&2C_{\mu}S^{\mu\alpha}\!-\!2B_{\mu}\Sigma^{\mu\alpha}\!=\!0\\
&C^{\alpha}\Phi\!+\!2qA_{\mu}\Sigma^{\mu\alpha}\!+\!(Y\Phi\!+\!m)V^{\alpha}\!=\!0\\
&B^{\alpha}\Phi\!-\!2qA_{\mu}S^{\mu\alpha}\!=\!0\\
&C_{\mu}V^{\mu}\!-\!qA_{\mu}P^{\mu}\!-\!(Y\Phi\!+\!m)\Phi\!=\!0\\
&C_{\mu}P^{\mu}\!-\!qA_{\mu}V^{\mu}\!=\!0\\
&B_{\mu}V^{\mu}\!=\!0\\
&B_{\mu}P^{\mu}\!=\!0
\end{eqnarray}
as conditions implicitly giving the maximal spinor field.

In the special case of neutral matter fields it is easy to assess that either the field has to be single-handed or alternatively we are left with the constraint
\begin{eqnarray}
&C^{\alpha}\Phi\!+\!(Y\Phi\!+\!m)V^{\alpha}\!=\!0
\label{n}
\end{eqnarray}
whose consequences will be discussed next.

So far we have moved in a rather general manner, but although the most general situation is the one that is richest with information, nevertheless it can turn out to be very instructive to study cases of special symmetry.

A special symmetry that is however important is the most general stationary axially symmetric background described by the Weyl-Lewis-Papapetrou metric, which is given in spherical coordinates $(t,r,\theta,\varphi)$ as
\begin{eqnarray}
&g_{tt}\!=\!e^{2\nu}\!-\!|Q\omega r\sin{\theta}|^{2}e^{-2\nu}\\
&g_{rr}\!=\!-e^{2(\lambda-\nu)}\\
&g_{t\varphi}\!=\!\omega|Qr\sin{\theta}|^{2}e^{-2\nu}\\
&g_{\theta\theta}\!=\!-r^{2}e^{2(\lambda-\nu)}\\
&g_{\varphi\varphi}\!=\!-|Qr\sin{\theta}|^{2}e^{-2\nu}
\end{eqnarray}
in which $\nu$, $\lambda$, $\omega$ and $Q$ are functions of the radial coordinate and elevation angle and all other elements are taken to be identically zero; it is a straightforward calculation to show that of all gravitational potentials only those given by $C_{r}$ and $C_{\theta}$ together with $B_{r}$ and $B_{\theta}$ are present with electrodynamic potentials $A_{t}$ and $A_{\varphi}$ and while all the other components are taken to vanish identically in all calculations we will perform. In this context, we have 
\begin{eqnarray}
&C_{1}\!=\!\frac{1}{4}e^{-(\lambda+\nu)}Q\sin{\theta}\partial_{\theta}\omega\\
&C_{2}\!=\!-\frac{1}{4}e^{-(\lambda+\nu)}Q\sin{\theta}r\partial_{r}\omega
\end{eqnarray}
and therefore in the case in which there is no relevant angular velocity $\omega\!\approx\!0$ with the consequence of the vanishing of the tidal potentials $C_{i}$ identically, while the gravitational potentials $B_{i}$ are given according to the expressions
\begin{eqnarray}
&B_{1}\!=\!\frac{1}{2}e^{(\nu-\lambda)}\partial_{r}\ln{|e^{(\lambda-\nu)}Qr^{2}\sin{\theta}|}\\
&B_{2}\!=\!\frac{1}{2r}e^{(\nu-\lambda)}\partial_{\theta}\ln{|e^{(\lambda-\nu)}Qr^{2}\sin{\theta}|}
\end{eqnarray}
and as already remarked in the case of maximal density the gravitational potentials $B_{i}$ are trivial in the sense that they can always be vanished by employing a mere coordinate transformation, and no assumption is made for the electrodynamic potential $A_{i}$ whatsoever.

In this situation, the damping factor might be integrated and the result is given by the form
\begin{eqnarray}
e^{-2\int\!B_{\nu}dx^{\nu}}\!\equiv\!\left|\frac{Ke^{(\nu-\lambda)}}{Qr^{2}\sin{\theta}}\right|
\end{eqnarray}
with $K$ integration constant, while integration of (\ref{ffe}) would give the behaviour of $\overline{\phi}\phi$ in general; according to this analysis, it is therefore possible to see that the condition given by the finiteness of the integral
\begin{eqnarray}
\left|\int^{\infty}_{0}\!\!\!\!\int^{\pi}_{0}\!\!\overline{\phi}\phi e^{(\lambda-\nu)}d\theta dr\right|<\infty
\label{condition}
\end{eqnarray}
specifies that the damping action of the gravitational field is more effective than the spreading tendency of the massive spinor field distribution, and as a consequence of this fact the confinement of the matter field takes place.

As we have done before in general, also in this case of special symmetry the condition of confinement facilitates studying the conditions of maximal density, because in this circumstance the gravitational potential becomes trivial as it can always be vanished by a simple choice of coordinates, and as a consequence the problem is reduced to the investigation of the single constraint given by
\begin{eqnarray}
qA^{\alpha}\Phi\!+\!(Y\Phi\!+\!m)P^{\alpha}\!=\!0
\end{eqnarray}
or equivalently in the form
\begin{eqnarray}
q^{2}A^{k}A_{k}\!=\!|m\!+\!Y\Phi|^{2}
\label{conditions}
\end{eqnarray}
which can be read as an implicit relationship among the points of the spacetime in which the field gets its maximal density value; because the gravitational potential is absent it is possible to choose the metric to be approximately flat, and as a further simplification we shall assume to work in a situation of almost perfect spherical symmetry, where the electrostatic potential is
\begin{eqnarray}
A_{t}\!=\!\frac{q}{4\pi}\frac{1}{R}
\end{eqnarray}
so that for the generic constant value $\Phi\!=\!m/V$ we get
\begin{eqnarray}
R\!=\!\frac{1}{4\pi}\frac{q^{2}}{m}\frac{V}{V\!+\!Y}
\label{formula}
\end{eqnarray}
which represents the equation defining the surface corresponding to the maximal density of the field, that is a sphere of radius (\ref{formula}): large values of the charge squared or small values of the mass have the effect of rendering large the radius, as it should be intuitive since the self-electric repulsion depends on $q^{2}$ while its own gravitational attraction depends on $m$ in general. Then, because the field $\Phi$ is usually interpreted as the inverse volume occupied by the field, smaller volumes means $V$ will have to be taken smaller, and correspondingly the radius will decrease; conversely larger volumes means $V$ will have to be taken larger, but the radius cannot increase more than the value $R_{\mathrm{max}}\!=\!q^{2}/4\pi m$ which is therefore to be considered as the maximal radius of the classical matter field distribution. For the classical electron, the maximal radius is given by $R_{\mathrm{max}}\!\approx\!10^{-15}$ meters, although it is normally obtained through quite different calculations.

We remark that in the case of neutral matter fields condition (\ref{n}) in this case reduces to $Y\Phi\!+\!m\!=\!0$ showing that for attractive potentials $Y$ is negative and the bi-linear spinor $\Phi\!=\!m/|Y|$ is a condensate solution while for repulsive potentials $Y$ is positive and the matter field has to be massless and single-handed necessarily.

It is important to remark that in this analysis the Dirac fields are considered regular. In recent years there has been a considerable work about a classification based precisely on the role of bi-linear spinor fields, which has shown that beside the usual regular Dirac field there are also singular Dirac fields known under the name of flag-dipoles or flag-poles, as it has been extensively discussed in references \cite{daSilva:2012wp,daRocha:2008we,Cavalcanti:2014wia, daRocha:2013qhu}: all these can be defined to be Dirac spinor fields because they are all solutions of the Dirac spinorial field equation, but one must care to discriminate what are the regular or the singular ones, and in the present paper only the Dirac fields that are regular have been considered. On the one hand, focusing only on regular Dirac matter field distributions might be a loss of generality, but on the other hand such assumption is necessary because interpreting the Dirac field as describing real matter distributions means forbidding Dirac fields with a vanishing bi-linear scalar. And this is so because the bi-linear scalar is interpreted as what describes the Dirac matter field distribution throughout the spacetime.

Obviously, the approximations we have used were too extreme to leave us confident that no property has been lost, and thus we cannot hope that this model had all of the features we would expect to find for extended matter field distributions; nevertheless, this model did incorporate at least some of the features one would expect to have for extended matter field distributions localized in a compact region of space: all these results are intriguing enough to be confident that more results will be found when more realistic cases will be studied, but on the other hand cases that are realistic enough not to become degenerate and at the same time special enough to provide a sufficient amount of information may be difficult to find, and it is likely that they can be obtained only when exact solutions will be found. Encouragingly, there have been investigations in which elementary particles were modelled in terms of extended matter field distributions with a definite shape such as a ring, a disk and a shell, as discussed in \cite{Carter:1968rr,Israel:1970kp} and \cite{l1}; also their stability properties have been investigated in \cite{l2}. Extensible distributions localized in compact regions give rise to a discrete mass spectrum in terms of which an entire family of particles may be described as different resonances of a single fundamental field, as it has been suggested by Dirac \cite{Dirac:1962iy}.

In none of the previously mentioned models have the effects of the non-linear spin-spin contact interactions been considered; here we did, although we have not quite reached the same amount of results: this circumstance is unfortunate but this is to be expected because non-linearities always produce a remarkable increase of the intrinsic difficulty met in the treatment of any problem.

Hopefully, more investigations might succeed in finding exact solutions, but the lack of integrability methods for non-linearly coupled fields makes the search difficult.
%%%%%%%%%%%%%%%%%%%%%%%%%%%%%%%%%%%%%%%%%%%%%%%%%%%%%%%%%%%%%%%%%%%%%%%%%%%%%%%%%%%%%%%%%%%%%%%%%%%
%%%%%%%%%%%%%%%%%%%%%%%%%%%%%%%%%%%%%%%%%%%%%%%%%%%%%%%%%%%%%%%%%%%%%%%%%%%%%%%%%%%%%%%%%%%%%%%%%%%
\section{Comments}
In the last part we have remarked upon the fact that because any given matter field has the tendency to spread over the entire spacetime, then a mechanism should be found to counterbalance this tendency, and have the extended matter field distribution localized in a compact region of the spacetime itself; this has been done after we have found the formal solution (\ref{solution}) displaying a factor that gives rise to a gravitationally-induced exponential damping which might balance the spreading tendency of the field distribution: for the most general stationary axially symmetric spacetime, we have discussed how the expression given by (\ref{condition}) is to be considered as the condition for the gravitationally-interacting extended matter to actually be confined. We have studied the locus of points where the spinor had maximal density and no gravitational interaction: the Weyl-Lewis-Papapetrou spacetime could be further approximated to a nearly Schwarzschild spacetime and the locus of maximal density was obtained from the balance between the action of the electrodynamic potential and the field itself and found to be nearly spherical, with radius of the same order of magnitude of the classical radius. Some remark has been given for the case of single-handed and massless neutral fields.

This model seems to point toward the fact that the extended matter field distribution localized in a compact region of spacetime should be interpreted as real, and as a consequence, we would like to comment on some features of this model and its character. A question we would like to address is whether it is possible to give a qualitative description of the shape of this extended distribution, and heuristically it is indeed reasonable that the distribution be described as a thin-shell: within the matter field equations, all the possible terms that are proportional to the matrix element $i\boldsymbol{\gamma}^{\mu}$ are the kinetic term and the gravitational potential justifying why the gravitational attraction may balance the spreading of the wave function, and nothing else could since there is no other contribution of that type; another type of contribution is the matrix element $\boldsymbol{\gamma}^{\mu}$ determining an interaction between tidal effects and electrodynamics with the non-linear potentials of spinor fields. The two effects are proportional to two different matrices and thus, acting on different components of the spinor, give rise to two complementary scales: the former contribution is determined only in terms of fundamental constants, so that its scale is to be proportional to the only fundamental length, which is the Planck length, while the latter contribution is determined in terms of the mass and charge of the particle and the torsion constant, so its scale is to be determined in terms of those and it cannot be assigned unless all constants are. The rough picture that emerges is that of a shell with thickness of the order of the Planck length and radius that in general is much larger and whose precise value depends on the particle itself.

Whether such a qualitative description could be made quantitative without going so far as to find all the exact solutions is something we do not know, and certainly more work should be done in this direction; one thing we have to stress however, is that in any case one may not leave gravitational effects aside, and therefore in a curved spacetime plane-waves cannot be possible, making impossible to implement quantization: in a situation in which we assume no quantization, we have to work out all the experimentally observed results in terms of physical processes of different nature. That this is possible was first proven by Schwinger by replacing quantization in terms of external sources for fields \cite{Schwinger:1989ix}; a different philosophy might consist in recovering the results usually obtained in terms of quantization precisely by exploiting the fact that quantization is impossible because of the non-linear potentials in the field equations. And that this can in fact be done has been first noticed and discussed in previous works \cite{Fabbri:2013gza} and \cite{Fabbri:2014naa} and deepened in the present paper.

It is fascinating that gravitational attraction could after all balance the spreading tendency of a matter field in order to have the extended field distribution localized in a compact region of spacetime, and it is even more intriguing that this situation might affect the dynamics of particles yielding effects analogous to those that are commonly ascribed to the presence of radiative processes.

But although finding exact solutions will bring more information, the search for exact solutions of such an extensively coupled system may be out of reach.
%%%%%%%%%%%%%%%%%%%%%%%%%%%%%%%%%%%%%%%%%%%%%%%%%%%%%%%%%%%%%%%%%%%%%%%%%%%%%%%%%%%%%%%%%%%%%%%%%%%
%%%%%%%%%%%%%%%%%%%%%%%%%%%%%%%%%%%%%%%%%%%%%%%%%%%%%%%%%%%%%%%%%%%%%%%%%%%%%%%%%%%%%%%%%%%%%%%%%%%
\section*{Conclusion}
In the present paper, we have started with a dramatic assumption: namely, that there would be no form of commutation relationships for the components of any operatorial re-definition of the fields, that is that there be no quantization of fields at all: this idea is actually not new, and we have recalled how it is conveniently used under the name of normal-ordering precisely in quantum field theory, and how it is systematically employed by Schwinger himself in his source field theory; we also discussed how such an assumption is compatible with the fact that there cannot be plane-wave solutions in presence of gravity, interpretable with the fact that there is no general way of quantizing fields in non-trivial backgrounds: as a consequence, one might think that it would fit just perfectly to recover the results usually obtained through quantization in terms of non-linear potentials since these are what renders quantization impossible, as suggested in reference \cite{Fabbri:2014naa} and here; on the other hand, we have specified that the parallel between quantum-like corrections and non-linear potentials will receive more supporting evidence when a more precise structure for the solution of the field distributions is eventually found.

Consequently, we have investigated such matter field distributions first in general terms, then by assuming more and more hypotheses and approximations in order to make the results clearer: we dealt with a general implicit solution of the form (\ref{mattersolution}); and we focused on the matter field equations studying the general properties in terms of formal solutions of the type (\ref{solution}): this last case was particularly instructive since it gave us the possibility to see that the solution can be confined by its own gravitational field to be localized into a compact region of space, and we have heuristically described how a thin-shell shape for the distribution might be obtained in the case of a stable configuration. In this heuristic description, the shell radius decreases with the mass and increases with the charge squared; for a given particle, the radius can be controlled by two constants of the problem, but there is a maximal radius that can be reached and it corresponds to the classical radius. Eventually, we have discuss what happens for neutral particles, either in the case of attractive or in that of repulsive potentials within the matter field equations, witnessing that such particles turn out to be constrained to be single-handed, again reaching a result common in quantum field theory.

As a final comment, we have stressed that although all these results are quite general nevertheless this is precisely the reason why we cannot compute the values of interesting quantities, unless exact solutions are found.

We have also stressed that the actual search for exact solutions might be too complicate for now.
%%%%%%%%%%%%%%%%%%%%%%%%%%%%%%%%%%%%%%%%%%%%%%%%%%%%%%%%%%%%%%%%%%%%%%%%%%%%%%%%%%%%%%%%%%%%%%%%%%%
%%%%%%%%%%%%%%%%%%%%%%%%%%%%%%%%%%%%%%%%%%%%%%%%%%%%%%%%%%%%%%%%%%%%%%%%%%%%%%%%%%%%%%%%%%%%%%%%%%%

%%%%%%%%%%%%%%%%%%%%%%%%%%%%%%%%%%%%%%%%%%%%%%%%%%%%%%%%%%%%%%%%%%%%%%%%%%%%%%%%%%%%%%%%%%%%%%%%%%%
\end{document}